\documentclass[nofootinbib,amsmath,amssymb,aps,prd,showpacs]{revtex4} 
\usepackage[colorlinks=true, pdfstartview=FitV, linkcolor=blue, citecolor=red, urlcolor=magenta]{hyperref}
\usepackage{graphicx,color}
\usepackage{latexsym}
\usepackage{amsmath}
\usepackage{amsfonts}
\usepackage{amssymb}
\newcommand{\be}{\begin{equation}}
\newcommand{\ee}{\end{equation}}
\newcommand{\ben}{\begin{eqnarray}}
\newcommand{\een}{\end{eqnarray}}
\usepackage{amsmath}
\usepackage{graphicx}
\newcommand{\sech}{\rm sech}

\begin{document}


\title{Braneworld solutions from scalar field in bimetric theory}

\author{D. Bazeia$^{a,b,c}$, F.A. Brito,$^{c}$ and F.G. Costa,$^{b,d}$}
\email{dbazeia@gmail.com; fabrito@df.ufcg.edu.br; geraldo.costa@ifrn.edu.br}
\affiliation{$^{a}$Instituto de  F\'\i sica, Universidade de S\~ao Paulo, 05314-970 S\~ao Paulo, SP Brazil\\
$^{b}$Departamento de  F\'\i sica, Universidade Federal da  Para\'\i ba, 58051-970 Jo\~ao Pessoa, PB Brazil\\
$^{c}$Departamento de F\'\i sica, Universidade Federal de Campina Grande, 58109-970 Campina Grande, PB Brazil\\ 
$^{d}$Instituto Federal de Educa\c c\~ao, Ci\^encia e Tecnologia do Rio Grande do Norte-IFRN, 59380-000 Currais Novos, RN Brazil}

\begin{abstract}
We investigate the presence of braneworld solutions in a bimetric theory, with gravity and the scalar field coupling differently. We consider a non-standard model, with a {\sl{Cuscuton-like}} scalar field, and we show how to generate braneworld solutions in this new scenario. In particular, we found no gravitational instabilities for the braneworld solutions.
\end{abstract}


\maketitle
\pretolerance10000


\section{Introduction}
The brane theory has been investigated as a candidate for solving the hierarchy problem, and other problems in high energy physics. In the Randall-Sundrum model \cite{lisa}, we can add scalar fields  with usual dynamics and allow them to interact with gravity in the standard way \cite{wise}. The study of scalar fields coupled to gravity in warped geometries has been frequently reported in the literature \cite{varios1,varios1a,varios2,varios3}, and in the current letter we consider a model driven by a single real scalar field. 

In recent years, there appeared some interesting models with non-canonical dynamics with focus on early time
inflation or dark energy, as good candidates to solve the coincidence problem \cite{picon1}. These kind of models have also been discussed in investigations of topological defects \cite{defeitos}. Basically, in these theories ones considers generalized dynamics, using in the Lagrange density a term in the form $F(X)$, with $X=\frac{1}{2}g^{ab}\partial_{a}\phi\partial_{b}\phi$.

In Ref. \cite{bazeia1} the generalized models are used in braneworld scenario with non-standard kinetic terms coupled with standard gravity and it is applied to a five-dimensional space-time to find new thick brane solutions.  Two specific examples of non-standard term were considered in \cite{bazeia1}: (I) $F(X)=X+\alpha|X|X$, and (II) $F(X)=-X^2$. In (I), in particular, one considers the case of small $\alpha$ and discuss perturbations around the case of a canonical scalar field due to the intrinsic nonlinear character of the Einstein equations, which usually results in a system of coupled ordinary differential equations that are very hard to solve.

On the other hand, a new class of actions with non-canonical kinetic terms has been used in Refs.~\cite{cusc1}, \cite{cusc2}. In the proposed theory, the equation of motion does not have the usual second order time derivative and the field becomes a non-dynamical auxiliary field, which plays the role of following the dynamics of the
fields that couple to it. For this reason, this field is known as {\sl{Cuscuton}}.

In the current work, we consider that the non-standard kinetic term arises due to the coupling with a different metric used to describe the gravitational field, that is, we use a \textit{bimetric} theory \cite{moffat}, \cite{drummond}, \cite{mag}. Bimetric theories have been proposed as VSL (varying speed of light) theories, motivated to solve the horizon, flatness, and dark matter problems. In the model proposed in \cite{moffat}, there are two metrics: $g_{\mu\nu}$ (which we refer to as the `gravitational metric') which is used to construct the Einstein-Hilbert action; and $\hat{g}_{\mu\nu}$ (which we refer
to as the `matter metric'), used to construct the matter action via minimal coupling, providing the geometry on which matter fields propagate and interact. Despite the simple structure of the equations of motion (in the case of a purely unconventional kinect term), such as the absence of second derivatives, we see that the model can describe brane solutions with a very simple potential. However, in this context, transformation connecting the two metrics has an additional term that comes from a vector normal to the brane. This term allows us to get to a $tachyonic$ action.

We organize the work as follows. In the next Sec.~\ref{bimetric} we look for flat thick 3-brane
solutions in a five-dimensional theory of gravity minimaly coupled with standard scalar field plus a five dimensional cosmological constant in bimetric theory generating a field with a non-conventional kinetic therm. In Sec.~\ref{first-order} we use the first order formalism \cite{varios1}, \cite{varios2}, \cite{varios3} to find solutions for some specific superpotentials. In Sec.~\ref{pure-cuscuton} we analyze the case in which the theory only contains the non-conventional term, the  (\textit{pure Cuscuton}) model, and we investigate localization
of gravity for the solutions we find. We end this work in Sec. \ref{concu}, where we include some conclusions and perspectives for future investigations.

\section{Bimetric and non-conventional dynamics for the scalar field}
\label{bimetric}

The non-standard braneworld scenario that we investigate is described by a theory of five-dimensional gravity coupled
to scalar and other matter fields governed by the action 
\be \label{action}S=S_{g}[g]+S_{\phi}[\phi, \partial_{a}\phi, g]+\hat{S}_{M}[\psi, \partial_{a}\psi, \hat{g}].\ee	
Here the gravitational action, $S_{g}$, is the Einstein-Hilbert action, constructed using the `Einstein' frame ($g_{ab}$) in the standard way
\be\label{agrav} S_{g}=-\frac{1}{2\kappa_5^2}\int d^{5}x\sqrt{\left|g\right|}\;R.\ee
The action for the scalar field is given by
\be S_{\phi}=\int d^{5}x\sqrt{g}\left[\;\eta\frac{1}{2}g^{ab}\partial_{a}\phi\partial_{b}\phi-V(\phi)\right],
\ee
where $\eta$ is a real parameter and $S_{M}[\psi, \hat{g}]$ is the matter action, where $\psi$ represents all the matter fields, with $\hat{g}$ being the metric on which the matter fields interact. To be explicit, we 
let the dynamics be driven by a bulk cosmological constant in the matter frame \cite{mag}
\be S_{M}[\psi, \hat{g}]=\int d^{5}x\sqrt{\left|\hat{g}\right|}\;\tilde{\Lambda}_{5}.\ee
Thus, the full action can be written as
\be\nonumber\label{full1} S=-\frac{1}{2\kappa_5^2}\int d^{5}x\sqrt{\left|g\right|}R+\int d^{5}x\sqrt{\left|g\right|}\left[\eta\frac{1}{2}g^{ab}\partial_{a}\phi\partial_{b}\phi-V(\phi)\right]+\int d^{5}x\sqrt{\left|\hat{g}\right|}\tilde{\Lambda}_{5}.\ee
For $\eta=1$ ($\eta=-1$) and $\tilde{\Lambda}_5=0$ we have a standard (phantom) scalar field theory coupled to gravity in the conventional way. For $\eta=0$ and $\tilde{\Lambda}_5\neq 0$ we have a pure {\sl{Cuscuton}} model in bimetric theory. 

The Ricci scalar is related to the Einstein frame ($g_{\mu\nu}$), while matter fields are coupled to the `matter metric'
where we mostly take $\kappa_5^2=2$, except as explicitly stated otherwise. Here $g_{ab}$ and $\hat{g}_{ab}$ describes the five dimensional spacetime, with $a, b = 0, 1, 2, 3, 4$ and $x_{4}\equiv y$ standing for the extra dimension.
The disformal transformation between the two metrics can be governed by a choice of dynamics. In the simplest case we can use a \textit{bi-scalar field} $\phi$ to write the relationship:
\be\label{hhg} \hat{g}_{ab}=g_{ab}+\epsilon B^{2}\partial_{a}\phi\partial_{b}\phi+Cu_{a}u_{b}.\ee
Here $u^{a}=(0, 0, 0, 0, 1)$ is a normal vector to the brane surface.  The main purpose of this decomposition 
in terms of the normal vector is to change dynamics in bimetric theory, such as DBI-like to {\sl{Cuscuton}} dynamics --- see below. One could also consider other decompositions. Another simple possibility is a four-dimensional cosmological scenario counterpart of this set up that could be achieved by making the coordinate time $t$ to develop the role of $y$ and assuming $u^{a}=(1, 0, 0, 0)$. Since $[\phi]=3/2$ then we have to have $[B]=-5/2$ with  $\epsilon$ and $C$ being both real dimensionless parameters. The line elements related to $g_{ab}$ and $\hat{g}_{ab}$ are given by
\be\label{met1} ds^2=e^{2A(y)}\eta_{\mu\nu}dx^{\mu}dx^{\nu}-dy^2,\ee
and
\be\label{met2} d\hat{s}^2=e^{2A(y)}\eta_{\mu\nu}dx^{\mu}dx^{\nu}-[(1-C)-\epsilon B^{2}\phi'^2]dy^2.\ee
Here we suppose that the scalar field only depends on the extra coordinate $y$, with the prime standing for derivative with respect to $y$. $e^{2A}$ is the warp factor, and $A = A(y)$ a real function of the extra dimension which gives rise to the warped
geometry. Also, $\eta_{\mu\nu}={\rm diag} (+---)$ describes the four-dimensional flat spacetime, with $\mu, \nu= 0, 1, 2, 3$. The
geometry of the five dimensional spacetime is then described by $A(y)$, and is driven by the extra coordinate $y$
alone. We can express the full metric in $g_{ab}$ framework. Using (\ref{met1}) and (\ref{met2}) we can write (\ref{full1}) as
\be\label{full2} S=-\frac{1}{4}\int d^{5}x\sqrt{\left|g\right|}R+\int d^{5}x\sqrt{\left|g\right|}\left[\;-\frac{\eta}{2}\phi'^2-V(\phi)\right]
+\int d^{5}x\sqrt{\left|g\right|}\sqrt{(1-C)-\epsilon B^2\phi'^2}\tilde{\Lambda}_{5}.\ee
Making $C=0$, $\epsilon=-1$ and $B=1$ the third term becomes a DBI-like action with a constant potential $\tilde{\Lambda}_{5}$. For $C=-1$, $\epsilon=-1$ and $B=1$ we have
\be\label{full3} S=-\frac{1}{4}\int d^{5}x\sqrt{\left|g\right|}R+\int d^{5}x\sqrt{\left|g\right|}\left[\;-\frac{\eta}{2}\phi'^2-V(\phi)\right]+\int d^{5}x\sqrt{\left|g\right|}\tilde{\Lambda}_{5} B\phi'.\ee
Here $L_{\phi}=-\frac{1}{2}\eta\phi'^2+\tilde{\Lambda}_{5} B\phi'-V(\phi)$ is an effective scalar field Lagrangian containing a non-standard kinetic term. The Lagrange density describing the scalar field can be written in the form $L_{\phi}=F(X, \phi)-V(\phi)$, with $F(X)=\frac{1}{2}X+\tilde{\Lambda}_{5}\sqrt{\left|X\right|}$ and $ X=g^{ab}\partial_{a}\phi\partial_{b}\phi$. Similar theory with time dependent scalar field was explored in the {\sl{Cuscuton}} cosmological model \cite{cusc1,cusc2}. Other more general models appear in braneworld scenario driven by scalar fields with non-standard kinetic terms coupled to standard gravity \cite{bazeia1}.


\section{Field equations and first order formalism}
\label{first-order}

Variation of (\ref{full3}) with respect to $g_{\mu\nu}$ leads (making $\eta=1$) to the equations:
\be \label{pri.deriv}A'^2=\frac{1}{6}\phi'^2-\frac{1}{3}V,\ee
and
\be \label{seg.deriv}A''=-\frac{2}{3}\phi'^{2}+\frac{2}{3}\tilde{\Lambda}_{5} B\phi'.\ee
The nonlinear character of the Einstein equations usually result in an intricate system of coupled ordinary differential equations that are hard to
solve. To find analytic solutions one can consider specific situations where first-order differential equations appear
describing the scalar field and metric functions, with the potential having a specific form \cite{bazeia1}, \cite{bazeia6}.

To get to the first-order formalism, we introduce the function, the $superpotential$ $W=W(\phi)$, which can be used to see the warp
factor as a function of the scalar field. We do this writing the first-order equation
\be\label{foeq1} A'=-\frac{1}{3}W.\ee
We use this equation and (\ref{seg.deriv}) to get to
\be\label{foeq2} \phi'=\frac{1}{2}W_{\phi}+\tilde{\Lambda}_{5} B,\ee
with the potential in (\ref{pri.deriv}) with the specific form
\be\label{pot1} V(\phi)=-\frac{1}{3}W^2+\frac{1}{2}\left[\frac{1}{2}W_{\phi}+\tilde{\Lambda}_{5} B\right]^2.\ee

The two equations \eqref{foeq1} and \eqref{foeq2} are the first-order differential equations we have to deal with to construct explicit solutions, for the potential
given by \eqref{pot1}. In the following we illustrate the procedure with two distinct examples.

\subsection{Flat brane solutions}

The first example is given by the well-known $\lambda\phi^4$ model obtained with \cite{bazeia6}
\be W(\phi)=2ab\left(\phi-\frac{b^2}{3}\phi^3\right).\ee
For this model one finds, for the scalar field 
\be\phi(y)=\sqrt{\frac{1}{b^2}+\frac{\tilde{\Lambda}_{5} B}{ab^3}}\tanh\left[ab^3\sqrt{\frac{1}{b^2}+\frac{\tilde{\Lambda}_{5} B}{ab^3}}y\right],\ee
and for the warp factor 
\be A(y)=\frac{1}{9}\left(\frac{2\tilde{\Lambda}_{5} B}{ab^{3}}-\frac{4}{b^2}\right)\ln\left[\cosh\left(ab^3\sqrt{\frac{1}{b^2}+\frac{\tilde{\Lambda}_{5} B}{ab^3}}y\right)\right] -\frac{1}{9}\left(\frac{\tilde{\Lambda}_{5} B}{ab^3}+\frac{1}{b^2}\right)\tanh^{2}\left(ab^3\sqrt{\frac{1}{b^2}+\frac{\tilde{\Lambda}_{5} B}{ab^3}}y\right).\ee
Let us now analyze the asymptotic behavior of the potential in the limits $y\rightarrow\pm \infty$. Here we have
\be V(\pm\infty)\equiv\Lambda_{5}=	-\frac{1}{3}\left(\frac{1}{b^2}+\frac{\tilde{\Lambda}_{5} B}{ab^3}\right) \left(\frac{4ab}{3}-\frac{2\tilde{\Lambda}_{5} B}{3}\right)^{2},\ee
where $\Lambda_{5}$ is an effective five dimensional cosmological constant. Note that for $\tilde\Lambda_5=2ab/B$ we have $\Lambda_{5}=0$, which corresponds to a $5D$ Minkowski ($M_{5}$) vacuum. For $\tilde\Lambda_5\neq 2ab/B$ ($ 1/b^2+\tilde{\Lambda}_{5} B/ab^3 >0$) we have $\Lambda_{5}<0$, that corresponds to an $AdS_{5}$ vacuum.

Another example is given by the superpotential \cite{bazeia1}
\be W(\phi)=3a\sin(b\phi).\ee
For this model we find the scalar kink profile
\be\label{Kink1} \phi(y)=\frac{2}{b}\arctan\left[\sqrt{\frac{2\tilde{\Lambda}_{5}B+3ab}{2\tilde{\Lambda}_{5}B-3ab}}\tan\left(\frac{b}{4}\sqrt{4\tilde{\Lambda}_{5}^2B^2-9a^2b^2}y\right)\right],\ee
and the warp factor 
\be\label{Warp1}
A(y)= -\frac2{3b^2}\ln\left[2\tilde{\Lambda}_{5}B+3ab-6ab\cos^2\left(\frac{b}{4}\sqrt{4\tilde{\Lambda}_{5}^2B^2-9a^2b^2}y\right)\right],
\ee
where we assume $\tilde{\Lambda}_{5}>{3ab}/{2B}$.
 In the limits $y\rightarrow \pm y^*$, where $y^*= 2\pi/b\sqrt{4\tilde{\Lambda}_{5}^2B^2-9a^2b^2}$, the potential in this case has the asymptotic behavior given by
\be\nonumber V(\pm y^*)\equiv{\Lambda}_{5}=\frac18(2\tilde{\Lambda}_{5} B-3ab)^2.
\ee
Following an analysis similar to that in the previous case we find that for $\tilde\Lambda_5=3ab/2B$ we have $\Lambda_{5}=0$, which would correspond to a $5D$ Minkowski ($M_{5}$) vacuum. However, this is not possible since the kink solution diverges in all space through this choice. On the other hand, for $\tilde\Lambda_5\neq 3ab/2B$ we have $\Lambda_{5}>0$, that corresponds to a $dS_{5}$ vacuum. This corresponds to an array of periodic kinks singular at $\pm y^*$ representing an array of braneworlds. 

Now assuming the $\tilde{\Lambda}_{5}<{3ab}/{2B}$ the solutions (\ref{Kink1}) and (\ref{Warp1}) become
\be\label{Kink2} \phi(y)=\frac{2}{b}\arctan\left[\sqrt{\frac{2\tilde{\Lambda}_{5}B+3ab}{|2\tilde{\Lambda}_{5}B-3ab|}}\tanh\left(\frac{b}{4}\sqrt{|4\tilde{\Lambda}_{5}^2B^2-9a^2b^2|}y\right)\right],\ee
and
\be\label{Warp2}
A(y)= -\frac2{3b^2}\ln\left[2\tilde{\Lambda}_{5}B+3ab-6ab\cosh^2\left(\frac{b}{4}\sqrt{|4\tilde{\Lambda}_{5}^2B^2-9a^2b^2|}y\right)\right].
\ee
We analyze the asymptotic behavior of the potential in the limits $y\rightarrow\pm \infty$ to get
\be
V(\pm \infty)\equiv{\Lambda}_{5}=-\frac1{3b^2}(3ab-2\tilde\Lambda_5B)(3ab+2\tilde\Lambda_5B).
\ee
Again as in the previous case we find that for $\tilde\Lambda_5=3ab/2B$ we have $\Lambda_{5}=0$, which would correspond to a $5D$ Minkowski ($M_{5}$) vacuum, but this choice is not possible, because the kink solution would diverge in all space. On the other hand, for $\tilde\Lambda_5\neq 3ab/2B$ we have $\Lambda_{5}<0$, that corresponds to an $AdS_{5}$ vacuum. Thus, the solution (\ref{Kink2}) corresponds to a regular kink representing a flat braneworld with geometry (\ref{Warp2}) embedded in a $AdS_{5}$ space. 

We see from the above results that in both examples, the part $\tilde\Lambda_5B$ related to the {\sl{Cuscuton}} tends to even out the $AdS_{5}$ and $dS_{5}$ spaces as $\tilde\Lambda_5B\to2ab$ or $\tilde\Lambda_5B\to3ab/2$. These results show that the {\sl{Cuscuton}} induces the tendency of a transmission of the gravitational dynamics to the scalar sector.  We shall further explore this effect for the pure Cuscuton dynamics in the next section.

\section{Solution with pure non-conventional dynamics}
\label{pure-cuscuton}

In this section we consider the case where the theory does not contain the conventional kinetic term (making $\eta=0$ in Eq. (\ref{full3})). So in this case the field equations are
\be\label{motion}4\tilde{\Lambda}_{5} A''=-\frac{\partial V(\phi)}{\partial\phi},\ee
\be \label{pri.deriv2}A'^2=-\frac{\kappa^{2}_{5}}{6}V,\ee
and
\be \label{seg.deriv2}A''=\frac{\kappa^{2}_{5}}{3}\tilde{\Lambda}_{5} B\phi'.\ee
Equations (\ref{pri.deriv2}) and (\ref{seg.deriv2}) lead naturally to the following result for the potential
\be V(\phi)=-\frac{2}{3}\kappa^{2}_{5}\tilde{\Lambda}_{5}^{2}B^{2}\phi^2,\ee
and (\ref{motion}) is a consistency equation. This is the only potential allowed in this case. However, in spite of its simplicity, we may find distinct solutions for the model.
Let us suppose that there is a \textit{kink-like} solution in the form
\be\label{campo} \phi(y)=b\tanh(ay).\ee
For this model the warp factor is
\be \label{warp}A(y)=-\frac{\kappa^{2}_{5}Bb\tilde{\Lambda}_{5}}{3a}\ln(\cosh(ay)).\ee
We note that the potential approaches the negative values $\Lambda_{5}\equiv V(\pm \infty)=-\frac{2}{3}\kappa^{2}_{5}\tilde{\Lambda}_{5}^{2}B^{2}b^{2}$ in the asymptotic limit $y\rightarrow \pm\infty $. This shows that the bulk is asymptotically $AdS_5$.

We can use the above results to construct the {\it thin-brane} limit for this solution, by taking the particular limit: $a\rightarrow \infty$ and $b\rightarrow 0$, with the product $ab$ fixed to a finite value. From (\ref{campo}) we see that 
\be \phi'(y)=ab\,\sech^{2}(\textit{ay}),\ee
such that in the thin-brane limit one obtains
\be \phi'(y)=ab\,\delta(y).\ee
We note that the presence of the delta function in (\ref{seg.deriv2}) and the absence of singularity in (\ref{pri.deriv2}) entail that $T^{0}_{0}=-\kappa^{2}_{5}\tilde{\Lambda}_{5}	 B\phi'+V\rightarrow-\kappa^{2}_{5}\tilde{\Lambda}_{5} Bab\,\delta(y)$ because $T^{5}_{5}=V=0$ in the limit $y\rightarrow0$. Thus, the effective brane tension becomes $\sigma=-\kappa^{2}_{5}\tilde{\Lambda}_{5} Bab$ and then we find that the solution (\ref{warp}) approaches 
\be A(y)=-\frac{\kappa^{2}_{5}}{3}\tilde{\Lambda}^{2}_{5}Bb\left|y\right|.\ee

\subsection{Gravity Localization}

 Let us now examine gravitational fluctuations in the above scenario. We do this perturbing the `gravitational metric', using
\be ds^2=e^{2A(y)}(\eta_{\mu\nu}+h_{\mu\nu})dx^{\mu}dx^{\nu}-dy^2.\ee
The wave function of graviton modes due to linearized gravity equation of motion in an arbitrary number of
dimensions ($d>3$) is given by \cite{varios1} 
\be\label{klein} \partial_{a}(\sqrt{\left|g\right|}g^{ab}\partial_{b}\Phi)=0.\ee
Let us consider $\Phi=h(y)\varphi(x^{\mu})$ and $\nabla^2_{4}\varphi=m^2\varphi$ into (\ref{klein}), where $\nabla^2_{4}\varphi$ is the flat four-dimensional Laplacian on the tangent frame. Thus the wave equation for the graviton through the transverse coordinate $y$ reads
\be\label{klein2} \frac{\partial_{y}(\sqrt{\left|g\right|}g^{yy}\partial_{y}h(y))}{\sqrt{\left|g\right|}}=m^2\left|g^{00}\right|h(y).\ee
This is our starting point to investigate both zero and massive gravitational modes on the 3-brane. Using the components of the metric (\ref{met1}) into the equation (\ref{klein2}) we have
\be\label{onda} \partial_{z}^2h+4\partial_{z}A\partial_{z}h=m^2e^{-2A}h.\ee

Now considering the following changes of variables: $h(y)=\psi(z)e^{-\frac{3A(z)}{2}}$ and $z(y)=\int e^{-A(y)}dy$ we can write (\ref{onda}) as a Schroedinger-like equation
\be\label{eqschr} -\partial_{z}^{2}\psi(z)+U(z)\psi(z)=m^2\psi(z),\ee
with a potential $U(z)$ given by
\be U(z)=\frac{3}{2}\partial^{2}_{z}A(z)+\frac{9}{4}(\partial_{z}A(z))^2.\ee
This equation can be factorized as
\be \left[-\frac{d}{dz}+\frac{3}{4}A'(z)\right]\left[\frac{d}{dz}+\frac{3}{4}A'(z)\right]\psi(z)=m^{2}\psi(z).\ee
So, there are no graviton bound-states with negative mass, and the graviton
zero mode $\psi_{0}(z)=e^{-3A(z)/2}$ is the ground-state of the quantum mechanical problem. Using (\ref{warp}) and making $a=\kappa^{2}_{5}\tilde{\Lambda}_{5}Bb/3$ 
we obtain
\be A(z)=\ln\left[\frac{1}{\sqrt{1+a^2z^2}}\right].\ee
The Schroedinger-like potential has the explicit form		
\be\label{schr} 
 U(z)=\frac{21a^4z^2}{4(1+a^2z^2)^2}-\frac{3a^2}{2(1+a^2z^2)}.\ee
Let us now investigate the  zero mode that corresponds to the solution of equation (\ref{eqschr}) with $m=0$. The general solution is given by
\be\label{fund} 
\psi_{0}(z)=\frac{a_{0}}{(1+a^2z^2)^{3/4}}+b_{0}\left[\frac{3\ln(az+\sqrt{1+a^2z^2})}{(1+a^2z^2)^{3/4}}+\frac{2a(5/2+a^2z^2)z}{(1+a^2z^2)^{1/4}}\right].\ee

In the brane scenario that we have just examined, in order for the zero mode to describe localized four dimensional gravity, normalizability is essential. To ensure normalizability, the zero mode as a function of $z$ must fall off faster than $z^{-1/2}$. To satisfy the normalization condition $\int^{+\infty}_{-\infty}\left|\psi_{0}(z)\right|^{2}=1$, we will consider  $a_{0}=\sqrt{a/2}$ and $b_{0}=0$ in (\ref{fund}) to obtain 
\be 
\psi_{0}(z)=\sqrt{\frac{a}{2}}\frac{1}{(1+a^2z^2)^{3/4}}.\ee
This is the zero mode that governs the localization of four-dimensional gravity on the brane.
We use $\Lambda_{5}=-\frac{2}{3}\kappa^{2}_{5}\tilde{\Lambda}_{5}^{2}B^{2}b^{2}$, as previously found, and the fact that $a=\kappa^{2}_{5}\tilde{\Lambda}_{5}Bb/3$ 
to find that $\Lambda_{5}\sim -{a^2}$; this means that $\sim 1/a$ is then related to the $AdS_5$ radius. This shows that the {\sl{Cuscuton}} fully controls the $AdS_5$
curvature, just as in conventional theories.

We end this section by commenting on the consistency of the whole theory. Since this braneworld set up is inspired in
its cosmological counterparts \cite{moffat,drummond,mag}, the
equation (\ref{hhg}) leads to a relationship among $g_{ab}$ and $\hat{g}_{ab}$. Thus, despite of our demonstration above, which shows that there is no gravitational instabilities, they may appear in the the scalar sector. However, the coupling $B$ in (\ref{hhg}) in general may be running with the scalar field \cite{Magueijo:2010zc} in such a way to compensate instabilities from the scalar field. In the present study we are simply assuming that $B\equiv B(\phi)$ runs to a very small constant.

\section{Conclusions}
\label{concu}

In this paper we connected two distinct and important issues addressed in high energy physics in relatively recent years. The first issue concerns the use of bimetric theories, whose main interest is to achieve realistic cosmological models; the other issue is the presence of generalized dynamics, such as the {\sl{Cuscuton}} model, whose dynamics itself is trivial, but when coupled to another sector of theory being gravitational or scalar, it may get dynamical behavior and then generate its own solutions. Here we also studied how to fine tune suitably the parameters of the theory, in order to turn the {\sl{Cuscuton}} model into the class of DBI-like theories. 

In addition, we have studied how the model behaves in the braneworld scenario. We found explicitly that the pure {\sl{Cuscuton}} sector acquires dynamics from the gravitational field to produce a braneworld solution which is able to localize four-dimensional gravity. As future investigations, one is now studying the construction of multiple {\sl{Cuscuton}} theories, which can be obtained from the presence of several scalar fields. Such studies should be compared with multi DBI-like theories, both in the braneworld and in the cosmological scenario, trying to understand how such distinct type of models evolve in the two scenarios.

\acknowledgments

We would like to thank CAPES, CNPq, FAPESP and PNPD/PROCAD-CAPES for partial financial support.
	

\end{document}